\documentclass[aps,prc,groupedaddress,showpacs,manuscript]{revtex4-1}
\usepackage{graphicx}

\usepackage{colordvi}

\begin{document}

\title{Influence of coalescence parameters on the production of protons and Helium-3 fragments}

\author {Qingfeng Li$\, ^{1}$\footnote{E-mail address: liqf@hutc.zj.cn},
Yongjia Wang$\, ^{1}$,
Xiaobao Wang$\, ^{1}$, and
Caiwan Shen$\, ^{1}$}

\affiliation{
1) School of Science, Huzhou University, Huzhou 313000, P.R. China \\
\\
 }
\date{\today}

\begin{abstract}

The time evolution of protons and $^3$He fragments from Au+Au/Pb+Pb reactions at 0.25, 2, and 20 GeV$/$nucleon is investigated with the potential version of the Ultrarelativistic Quantum Molecular Dynamics (UrQMD) model combined with the traditional coalescence afterburner. In the coalescence process, the relative distance $R_0$ and relative momentum $P_0$ are surveyed in the range of 3-4 fm and 0.25-0.35 GeV$/$c, respectively. For both clusters, a strong reversed correlation between $R_0$ and $P_0$ is seen and it is time-dependent as well. For protons, the accepted ($R_0$, $P_0$) bands lie in the time interval 30-60 fm$/$c, while for $^3$He, a longer time evolution (at about 60-90 fm$/$c) is needed. Otherwise, much smaller $R_0$ and $P_0$ values should be chosen. If we further look at the rapidity distributions from both central and semi-central collisions, it is found that the accepted [$t_{\rm cut}, (R_0, P_0$)] assemble can provide consistent results for proton yield and collective flows especially at mid-rapdities, while for $^3$He, the consistency is destroyed at both middle and projectile-target rapidities.

\end{abstract}



\maketitle

\section{motiviation}
The production mechanism of particles and nuclei is a fundamental and essential problem for the whole evolution of the universe. With the help of macroscopic and/or microscopic transport models for heavy ion collisions (HICs) within a large range of beam energies from the GSI Schwerionen Synchrotron (SIS) up to the BNL Relativistic Heavy Ion Collider (RHIC) and the CERN Large Hadron Collider (LHC) energies, the process of a complete compression and decompression from initial two colliding nuclei can be described dynamically, with the possible occurrence of (phase) transitions from nuclear liquid to gas (LG) or from quark-gluon plasma (QGP) to hadron gas (HG). Such a model is, e.g. the Ultrarelativistic Quantum Molecular Dynamics (UrQMD)\cite{Bass98,Bleicher99} \footnote{see UrQMD homepage, www.urqmd.org.} . After that, an afterburner is usually chosen for the freezeout of various particles (free baryons, mesons, etc.) or fragments (deuterons, tritons, Helium isotopes, etc.) which can then be used for comparison with corresponding experimental data. It is unavoidable that the afterburner should be paid much more attention when a serious comparison between calculations and experiments is needed for extracting, although, mainly the information of the compression phase at the early stage. Therefore, a large amount of such models are available but active in different beam energy regions (with their own problems when describing data). At low bombarding energies from several tens to several hundreds MeV$/$nucleon, the statistical multifragmentation model (SMM) \cite{Bondorf:1985mv,Bondorf:1995ua}, the statistical evaporation model (HIVAP) \cite{Weisdor:1981xx,Weisdor:1985xx} and the statistical model (GEMINI) \cite{Charity:1988zz} are frequently used; at energies from several hundreds MeV$/$nucleon to several GeV$/$nucleon, the conventional phase-space coalescence model ``Minimum Spanning Tree'' (MST) \cite{Kruse:1985pg} as well as the Simulated Annealing Clusterisation Algorithm (SACA) \cite{Gossiaux:1997hp} are taken into account; at higher energies, the coalescence afterburner using a Wigner-function method \cite{Mattiello:1995xg,Nagle:1996vp,Monreal:1999mv} is ordinarily in use.

Recently, the SMM has been extended and successfully combined with the Giessen Boltzmann-Uehling- Uhlenbeck (GiBUU) transport model in order to describe future experiments in hypernuclear physics at, e.g. the new GSI- and J-PARC-facilities \cite{Gaitanos:2009at}
. However, the hyperfragments are still constructed within the phase space coalescence model. Further, a so-called surface coalescence mechanism was introduced into QMD-like models such as the JAERI QMD (JQMD) \cite{Watanabe:2007xx} and the Improved Quantum Molecular Dynamics (ImQMD) \cite{Wei:2013wfa} in order for a better description on the experimental data of light complex particles (LCP) produced in nucleon-induced spallation reactions. Quite recently, the traditional coalescence afterburner has been combined with a potential version of UrQMD model to explain the production of free protons and $^3$He clusters from HICs at the Brookhaven Alternating Gradient Synchroton (AGS) and the CERN Super Proton Synchrotron (SPS) energies \cite{Li:2016mqd,Li:2016xx} and it was found that, with only one parameter set of ($R_0$,$P_0$)=(3.8 fm, 0.3 GeV$/$c) but using different stopping times for the former dynamic process, the rapidity distribution of both yields can be described reasonably well.

Hence, it seems that the traditional phase-space coalescence method can be widely used for ion collisions in a large range of beam energies despite of its simplicity and defects in other aspects such as the binding energy and the isospin. As a benchmark test, it is thus interesting to study the production of free protons and $^3$He from HICs with beam energies from several hundreds MeV$/$nucleon to several tens GeV$/$nucleon (i.e., within two-order-of-magnitude energies). Further, due to the obvious kinetics/dynamics during collisions, there exists a strong correlation between coordinate and momentum spaces of particles, and this correlation is certainly time dependent as well. Therefore, it is necessary to make a scan of the time evolution of the parameter set ($R_0$,$P_0$) in a reasonable range for the coalescence process to produce the same yield of a certain particle. Based on these tests, the influence of coalescence parameters on observables related to protons and $^3$He clusters other than their yields can then be observed.

The paper is arranged as follows. In the next section, the potential version of the UrQMD and the choice of parameters $R_0$ and $P_0$ used in the traditional coalescence model are introduced briefly. In sect.3, the time evolution of free proton and $^3$He yields from central HICs at three beam-energy points lying in SIS, AGS, and SPS regions, respectively, are scanned and discussed. Their rapidity distributions produced from semi-central HICs are also investigated for understanding the influence of the afterburner on collective flows, such as the directed and elliptic ones, which are frequently used to extract comprehensive information on the (isospin-dependent) equation of state (EOS). Finally, a summary and outlook is given in sect. 4.

\section{the UrQMD and the coalescence model settings}

The UrQMD microscopic transport model was originally developed to study particle production at high energies such as AGS, SPS, and RHIC energies \cite{Bass98,Bleicher99,Bratkovskaya:2004kv}. Recently, it has been updated for simulating HICs at both lower, such as SIS energies \cite{Li:2011zzp,Guo:2012aa,Wang:2012sy,Wang:2013wca,Wang:2014aba} and higher, such as LHC energies \cite{Li:2012ta,Graef:2012za,Graef:2012sh}. It is interesting to see that the potentials always play an important role on the particle emission from HICs at whatever low or high energies. The current potential version of the UrQMD (UrQMD/M) has considered mean-field potentials for both formed baryons and pre-formed hadrons in a similar way and expressed as\cite{Li:2007yd}:

\begin{equation}
U(\rho_h/\rho_0)=\mu (\frac{\rho_h}{\rho_0})+\nu
(\frac{\rho_h}{\rho_0})^g, \label{den3}
\end{equation}
where $\mu$, $\nu$ and $g$ are parameters which determine the stiffness
of the EoS of nuclear matter (in this work, the same soft EoS with momentum dependence is adopted and the incompressibility $K=314$ MeV). $\rho_0$ is the normal nuclear density and $\rho_h$ is the density for both pre-formed hadrons and formed baryons (including anti-baryons). For formed mesons, no nuclear potential is considered as before. Certainly, the quark number difference between pre-formed baryons and mesons, and the relativistic effect on relative distances and momenta of two particles have been taken into account in the corresponding potential modification process. Then, the required information of all particles is poured into the coalescence model for each 10 fm$/$c time step before the final stopping time $t_{\rm cut}=100$ fm$/$c. Certainly, it is known that, with a large effort in constraining the stiffness of the EoS of isospin symmetric nuclear matter \cite{Danielewicz:2002pu}, a soft EoS with the incompressibility $K_0=230\pm30$ MeV has been approved extensively \cite{Dutra:2012mb}. Nevertheless, large uncertainties still exist in the isospin-dependent part of EoS especially at supranormal densities \cite{Xiao:2008vm,Feng:2009am,Russotto:2011hq}. And, the main conclusions drawn in this paper will not be changed by the current choice of the stiffness of EoS.

At each $t_{\rm cut}$ from 10 fm$/$c to 100 fm$/$c, the nucleons with relative momenta $\delta p<P_0$ and relative distances $\delta r<R_0$ will be considered to belong to one cluster. Certainly, baryons other than nucleons can be treated in a similar way but not discussed in this work. As a matter of experience, the values for the parameter set ($R_0$, $P_0$) might be chosen in the range of (3-4 fm, 0.25-0.35 GeV$/$c) in order to describe experimental data from HICs at SIS, AGS, and even SPS energies. Within such a large beam energy region, the relativistic effect on $\delta r$ and $\delta p$ between two baryons  have been taken into account by the well-known Lorentz transformation (LT) from the computational two-nucleus center-of-mass system to the local rest frame of two particles \cite{Li:2016mqd,Li:2016xx}, and it is found that it has visible effect on light fragments, although not on free nucleons, which should be paid attention.

At the early stage of the reaction, a large amount of constructed light clusters will be emitted pre-equilibrately, which should be constrained by existing experimental data. In this paper we select the data from three reactions: free protons from both (I) central (with $b/b_0<0.15$ where $b_0$ the sum of the radii of the colliding nuclei) Au+Au at $E_b=0.25$ GeV$/$nucleon \cite{Reisdorf:2010aa} and (II) central (with $\sigma/\sigma_T<5\%$ where $\sigma_T$ the total cross section) Au+Au at $E_b=2$ GeV$/$nucleon \cite{Klay:2001tf}, and free protons and $^3$He from (III) central (with $\sigma/\sigma_T<5\%$) Pb+Pb at $E_b=20$ GeV$/$nucleon \cite{Blume:2007kw,Strobele:2009nq}. In the reaction (I), the total number of free protons is about $34\pm8$ in which a large error bar exists. In the reaction (II) and (III), the total free protons can be extracted (non-directly) from their  rapidity distributions and are about $116\pm2$ and $145\pm1$, the small error bars are just a rough estimate for further use. Based on the same idea, the number of $^3$He fragments from the reaction (III) within the rapidity region $|y|<1.7$ (where $y=\frac{1}{2}\mathrm{log}(\frac{E_{\mathrm{cm}}+p_{//}}{E_{\mathrm{cm}}-p_{//}})$, and $E_{\mathrm{cm}}$ and $p_{//}$ are the energy and longitudinal momentum of the observed particle in the center-of-mass system, respectively) is taken as $0.22\pm0.02$.

\begin{figure}[htbp]
\centering
\includegraphics[angle=0,width=0.9\textwidth]{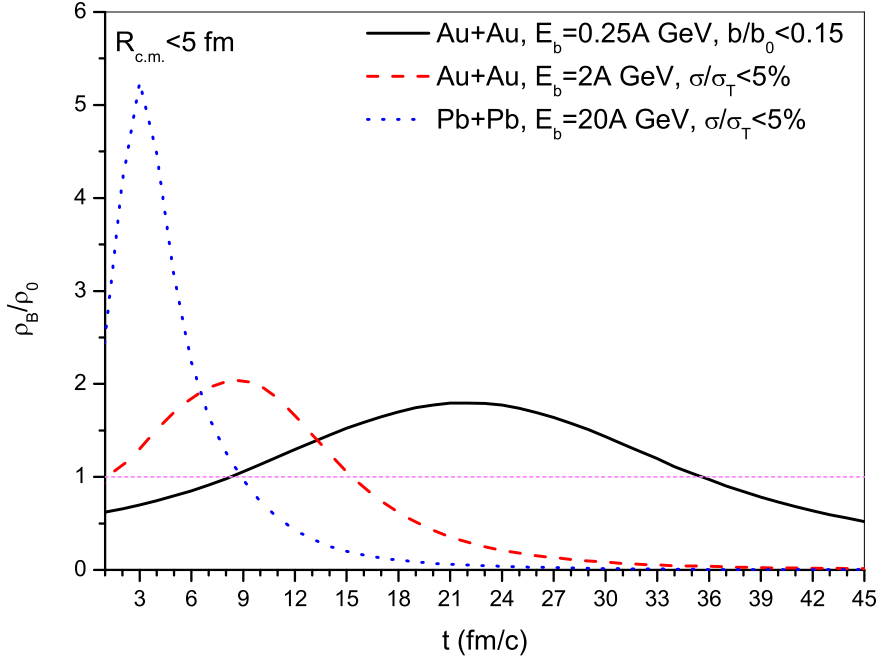}
\caption{\label{fig1} (Color) Time evolution of reduced densities of all baryons ($\rho_B/\rho_0$) in the central zone ($R_{\rm c.m.}<5$ fm) of the three reactions shown by different lines. The horizontal short-dashed line represents the normal density.
}
\end{figure}

Fig.~\ref{fig1} depicts firstly the time evolution (before 45 fm$/$c) of reduced densities of all baryons ($\rho_B/\rho_0$) in the central zone ($R_{\rm c.m.}<5$ fm) of the three reactions (shown by different lines). The horizontal short-dashed line represents the normal density. It is seen obviously that the maximum central density is enhanced largely when the beam energy increases from SIS to SPS. At $0.25$ GeV$/$nucleon, it is less than 2 times of the normal density, while at $20$ GeV$/$nucleon it becomes 5 times more than the normal density where the phase transition from QGP to HG might occurs. Meanwhile, the time interval for persisting $\rho_B/\rho_0>1$ is shortened largely from about 25 fm$/$c to 9 fm$/$c with increasing beam energy. With such a large change in central densities, it is curious to see how it influences the production of particles and fragments in the end of evolution.

\section{Time evolution of the production of protons and $^3$He}

\begin{figure}[htbp]
\centering
\includegraphics[angle=0,width=0.9\textwidth]{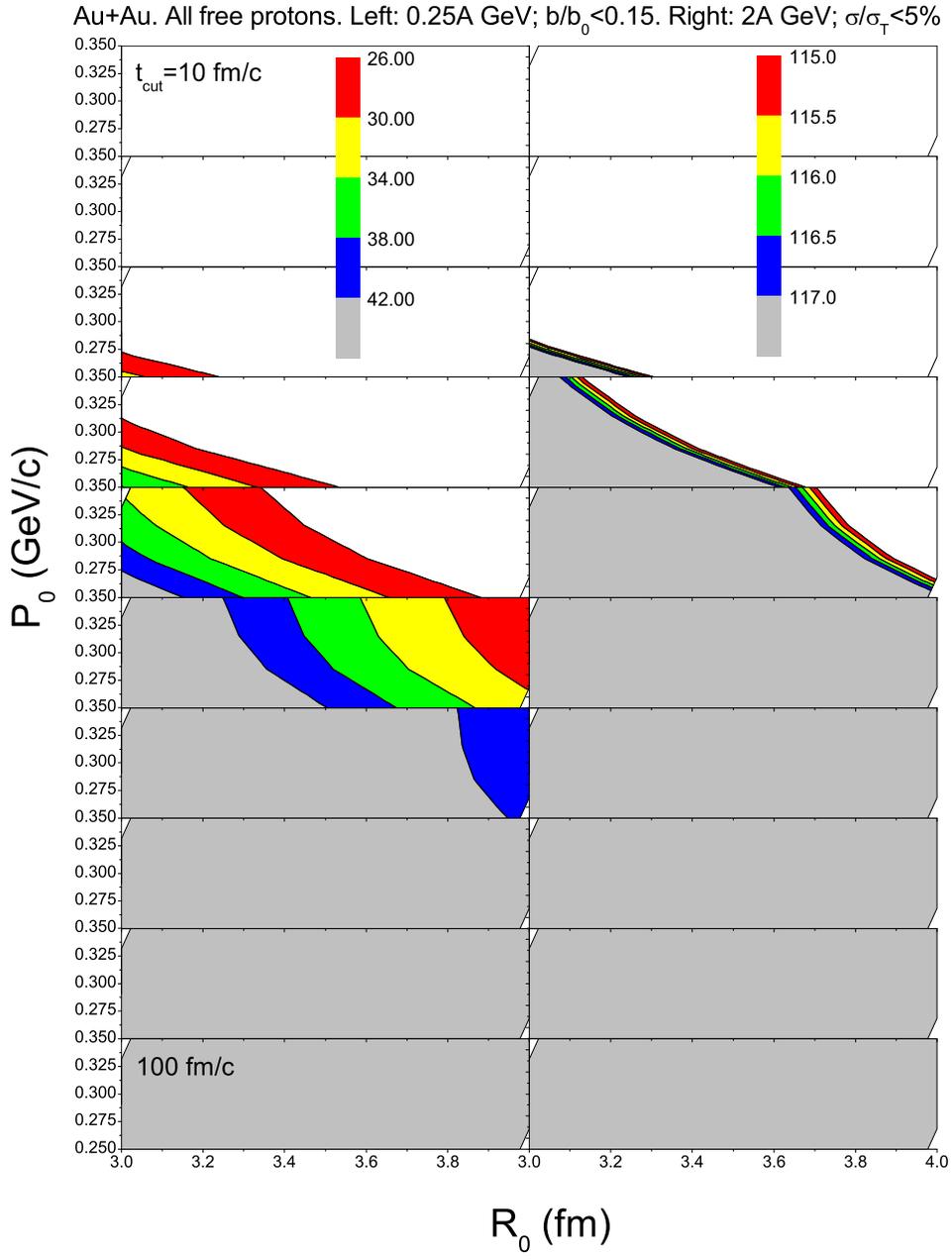}
\caption{\label{fig2} (Color) Time evolution of $R_0-P_0$ contour plots of free protons from the reaction I (left) and II (right).
}
\end{figure}

Fig.~\ref{fig2} shows the time evolution (from top to bottom for the time evolving from 10 fm$/c$ to 100 fm$/$c and with the time interval 10 fm$/$c) of contour plots for free proton yields from the reaction I (left) and II (right). The parameters $R_0$ and $P_0$ in the coalescence model range from 3-4 fm, and 0.25-0.35 GeV$/$c, respectively. In each plot, the colorful region is allowable for describing the corresponding experimental data, while the grey and/or white areas are out of that region. In the viewed ranges of $R_0$ and $R_0$, it is clear that enough protons are produced only after 30 fm$/$c and excessive yields happen after 70 fm$/$c. During this time period, the accepted area demonstrates a strong reversed correlation between $R_0$ and $P_0$ which is easy to understand due to the strong coupling nature of the dynamic process. If the errors are smaller, the $R_0-P_0$ correlation will be more certain. Furthermore, it is found that the correlation is time-dependent and the absolute value of the slope becomes larger with increasing time, which means that the proton production depends more heavily on the choice of $R_0$ than $P_0$. It is due to the fact that the influence of both mean-field potentials and two-body collisions becomes weaker at later times and the change in the particle momentum is much less. In addition, it is nice to see that at the same time of 50 fm$/$c, the yields of protons at both bombarding energies can be described well with the coalescence model, although values of the ($R_0$,  $P_0$) set should be larger at the higher beam energy.

\begin{figure}[htbp]
\centering
\includegraphics[angle=0,width=0.9\textwidth]{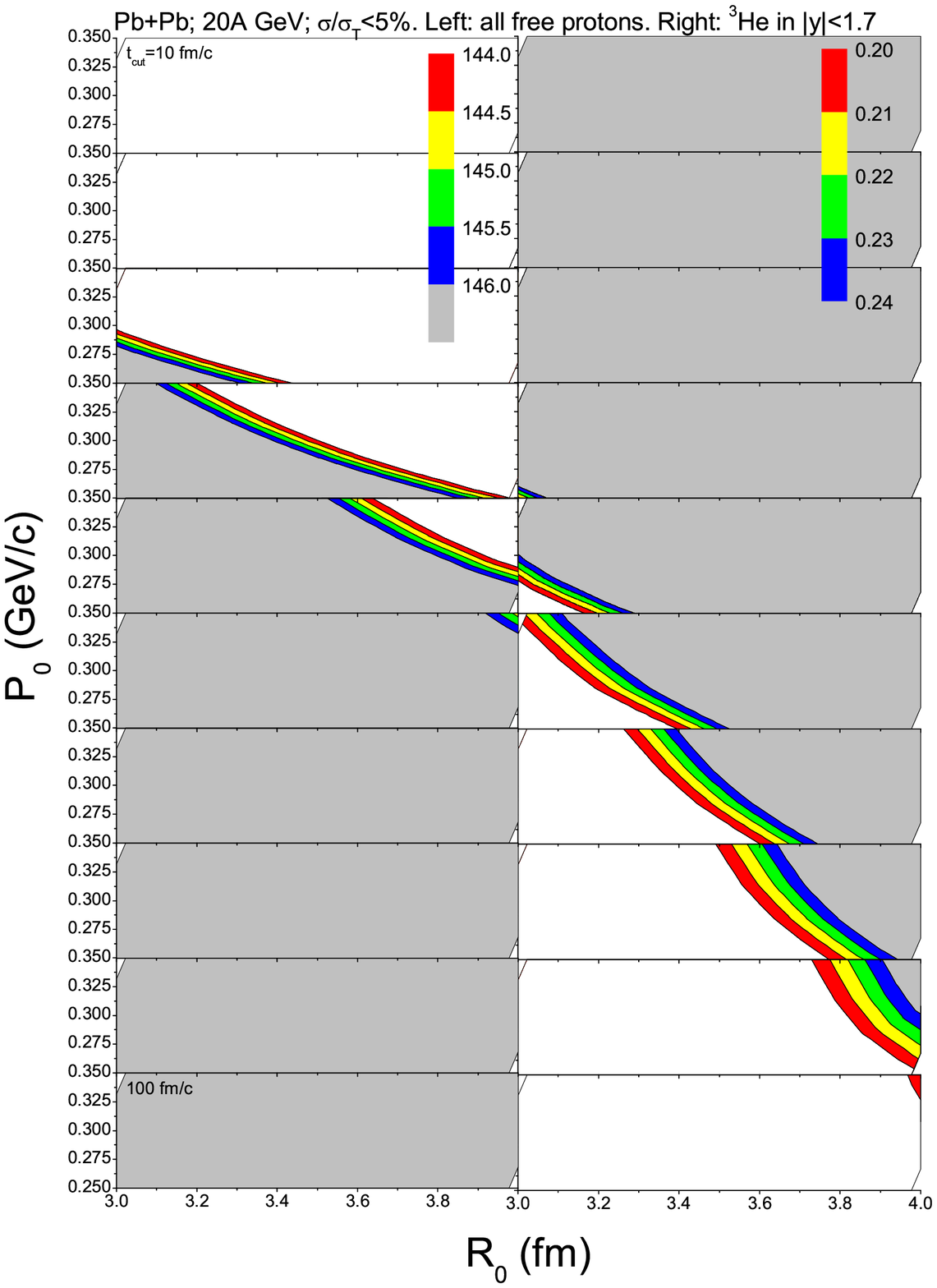}
\caption{\label{fig3} (Color) Time evolution for free protons (left) and $^3$He (right) from the reaction III.
}
\end{figure}

In the left plots of Fig.~\ref{fig3} we further show the time evolution for free proton yields from the reaction III. It is also found that the accepted time period for describing data lies in 30-60 fm$/$c, and, the corresponding $R_0-P_0$ correlation is quite similar to that at $E_b=2$ GeV$/$nucleon. All above results in the contour plots imply that the protons are produced at $\sim 50$ fm$/$c and weakly depend on the beam energy covering SIS, AGS, and SPS. However, at such an early time, the calculated yield of $^3$He can hardly meet the data, which is shown in the right plot of Fig.~\ref{fig3}: before 50 fm$/$c, too many light clusters such as $^3$He are produced but not stable and need time to decay. If we select two parameter sets of ($R_0$,  $P_0$), (3.5 fm, 0.3 GeV$/$c) and (3.8 fm, 0.3 GeV$/$c), which can both describe data well, it is found that the stopping times should be at 70 and 90 fm$/$c, respectively. This phenomenon was also seen in the previous work \cite{Li:2016xx} even for higher SPS energies.

By now, it is clear that the yields of both protons and $^3$He from HICs at energies ranging from SIS up to SPS can be described well within the cooperated UrQMD+coalescence approach if one proper assemble of the stopping time for UrQMD and a parameter set of ($R_0$,  $P_0$) in the coalescence is chosen. Further, the strong $R_0-P_0$ correlation provides us with a series of parameter sets of ($R_0$,  $P_0$) which helps us to check other observables which might be influenced by the final stage.

\begin{figure}[htbp]
\centering
\includegraphics[angle=0,width=0.9\textwidth]{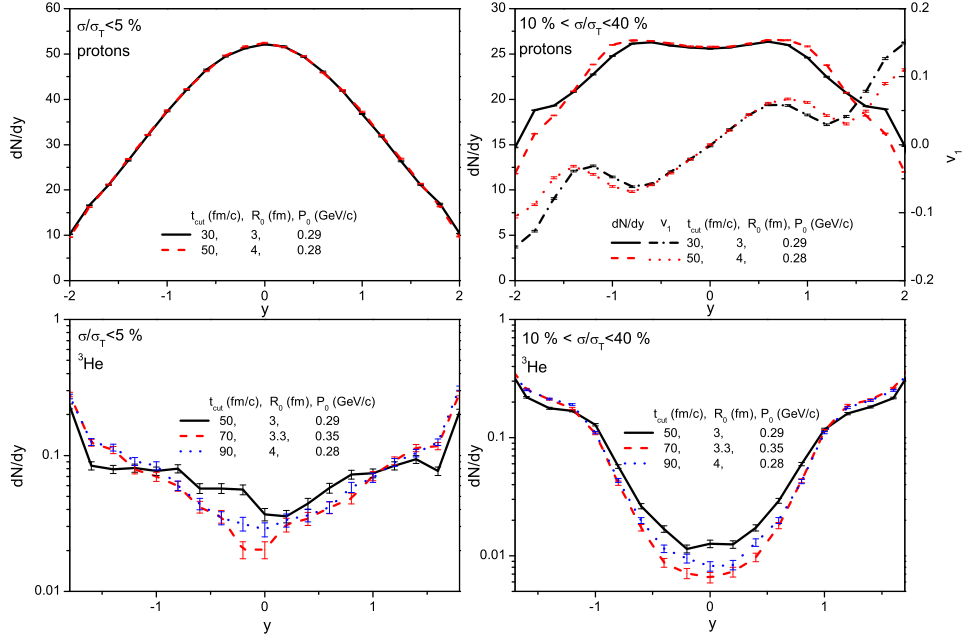}
\caption{\label{fig4} (Color) Rapidity distribution of proton (top plots) and $^3$He (bottom plots) yields from both central ($\sigma/\sigma_T<5\%$, left plots) and semi-central ($10\%<\sigma/\sigma_T<40\%$, right plots) Pb+Pb reactions at 20 GeV$/$nucleon. Different [$t_{\rm cut}, (R_0, P_0$)] parameter sets are used to demonstrate the largest effect on their rapidity distributions. In the top-right plot, the right axis is taken to shown the corresponding $v_1$ values.
}
\end{figure}

In the top two plots of Fig.~\ref{fig4} we firstly present the rapidity distribution of proton yields (using the left axis) from both central ($\sigma/\sigma_T<5\%$, left plot) and semi-central ($10\%<\sigma/\sigma_T<40\%$, right plot) Pb+Pb reactions at 20 GeV$/$nucleon. Two sets of [$t_{\rm cut}, (R_0, P_0$)] parameters are chosen to demonstrate the largest effect on the rapidity distribution, which are shown by solid and dashed lines. Besides the total yield, the rapidity distribution of the proton yield has almost no difference for central collisions as seen in the top-left plot. While for non-central collisions shown in the top-right plot, this situation has changed to some extend outside of the mid-rapidity region $|y|<0.6$. On the one side, the scanning work for values of $R_0$ and $P_0$ lying in the colorful regions of Figs.~\ref{fig2} and \ref{fig3} are useful since they are reliable in the description of proton data in total and in the mid-rapidity region. On the other side, the consistency check of the proton yields with the two different parameter sets guarantees the invariance of collective flow parameters $v_1$ and $v_2$ at mid-rapidities. As an example, in the top-right plot and taking the right axis into use, the corresponding $v_1$ values are exhibited with dash-dotted and dotted lines.

Then, in the bottom plots of Fig.~\ref{fig4} the rapidity distribution of $^3$He yields from the same central (left) and semi-central (right) Pb+Pb reactions as that of protons are shown, but with different [$t_{\rm cut}, (R_0, P_0$)] parameter sets (shown by lines). For central collisions, although the rapidity-integrated (in $|y|<1.7$) $^3$He yields are the same to each other, the rapidity distributions with different parameter sets are visibly different at both mid-rapidities and projectile-target rapidities. For semi-central collisions, this difference also exists. Meanwhile, the flow parameters $v_1$ and $v_2$ of $^3$He even at mid-rapidities are found to be influenced as well which should be paid attention.

\section{Summary and Outlook}
To summarize, the time evolution of emitted protons and $^3$He fragments from HICs at SIS, AGS, and SPS energies (represented by three energy points 0.25, 2, and 20 GeV$/$nucleon) is investigated with the potential version of UrQMD (UrQMD/M) combined with the traditional coalescence afterburner. And, in the coalescence process, the values of the ($R_0$, $P_0$) parameter set are surveyed in reasonable ranges (3-4 fm, 0.25-0.35 GeV$/$c) in order to describe experimental data of yields of both protons (in total) and $^3$He (in the rapidity region $|y|<1.7$) from central collisions. For both clusters, a strong reversed correlation between $R_0$ and $P_0$ values is seen and it is time-dependent as well. For protons, the accepted ($R_0$, $P_0$) bands lie in the time period 30-60 fm$/$c, while for $^3$He, a longer time evolution (at about 60-90 fm$/$c) is needed. Otherwise, much smaller $R_0$ and $P_0$ values should be chosen. When looking into their rapidity distributions from both central and semi-central collisions, it is found that the accepted [$t_{\rm cut}, (R_0, P_0$)] assemble can provide consistent result for proton yield and collective flows especially at mid-rapdities, while for $^3$He, the consistency is destroyed at both mid-rapidities and projectile-target rapidities, which deserves more investigation.

An obvious and easy-to-operate progress is to further consider the binding-energy check for the coalescence process, which has been examined for Au+Au collisions at the ALADIN energies recently \cite{Goyal:2011zz} and the predictive power is improved. With a further consideration of the energy contribution from momentum-dependent and symmetry potentials for the calculation of total binding energy, some isospin-sensitive observables related to light clusters were found to be obviously influenced \cite{Kumar:2015dfa}. Although uncertainties in determining (un-)stable fragments with large isospin-asymmetry exist, especially when the beam energy rises to AGS and higher, the more realistic binding-energy check is necessary and becomes more feasible when the large ($R_0$, $P_0$) bands at various stopping times for the earlier dynamic transport process are available for further constraints.

\begin{acknowledgements}
We acknowledge support by the computing server C3S2 in Huzhou
University. The work is supported in part by the National
Natural Science Foundation of China (Nos. 11375062, 11547312, 11275068, 11505056, 11505057), the project sponsored by SRF for ROCS, SEM, and the Doctoral Scientific Research Foundation (No. 11447109).
\end{acknowledgements}

\end{document}